\documentclass[twoside,11pt]{article}

\usepackage{graphicx}

\usepackage{amssymb}
\usepackage{amsmath}
\usepackage{amsfonts}
\usepackage{wasysym}
\usepackage{revsymb}
\usepackage{amsbsy}
\usepackage{bm}
\usepackage{MnSymbol}
\usepackage{upgreek} 

\usepackage{titlesec}

\usepackage{cite}

\titleformat*{\section}{\bfseries}
\titleformat*{\subsection}{\bfseries}

\begin{document}           

\title{\Large 
Scalar fields with power-law potentials in quantum cosmology}
\author{V.E. Kuzmichev, V.V. Kuzmichev\\[0.5cm]
\itshape Bogolyubov Institute for Theoretical Physics,\\
\itshape National Academy of Sciences of Ukraine, Kyiv, 03143 Ukraine}

\date{}

\maketitle

\begin{abstract}
A homogeneous and isotropic quantum cosmological system (universe) initially filled with a uniform scalar field that has a potential in the power law 
representation is considered. Depending on the epoch, this scalar field yields barotropic matter in the form of stiff matter, perfect gas, radiation, dust, cosmic 
strings, domain walls, de Sitter vacuum, or phantom matter. The proposed approach is based on quantum geometrodynamics for the maximally symmetric 
space. The relevant differential equations for the separate power-law summands of the scalar field potential and the corresponding quantum Hamiltonian 
constraint equations, which describe the universe that can be viewed as dominated by one form or another of barotropic matter, were obtained. The solutions 
to these equations have been found in analytical form.
\end{abstract}


\section{Introduction}\label{sec:1}
Scalar fields play an essential role in modern cosmology. They allow one to explain the hypothesized inflationary stage of the early universe's evolution, and, 
with their help, one can simulate dark matter and dark energy in the late universe. Scalar fields can be viewed more broadly as an alternative framework for 
describing the matter content of the universe. Instead of describing matter from the beginning as the cosmological fluid in terms of its energy density and 
pressure, it is possible to introduce the scalar field which serves as a surrogate for the matter content of the universe. 

Models involving the standard types of cosmological matter have been studied in the literature (for details, see, e.g. 
Refs.~\cite{CG82,DS86,D96,CG01,DS06}). In classical cosmology, matter content is often described in terms of a perfect fluid that satisfies a barotropic 
equation of state $p = p(\rho)$, where $p$ is the pressure and $\rho$ is the energy density of the matter. However, in certain cases, it is thought to be more 
appropriate to describe the matter content of the universe using scalar field models than to consider it in terms of fluids \cite{SHS14}. As a matter of fact, these 
two approaches to representing matter-energy are closely related. It was found that the energy momentum tensor of a scalar field has the same 
structure as a fluid energy momentum tensor, even though scalar fields and perfect fluids are quite different systems from a physical point of view 
\cite{M85,M88,F12, FGGV23}. Specifically, there is an analogy between a scalar field minimally coupled to gravity and a perfect fluid. For this reason, scalar 
field models can mimic the standard cosmological models with perfect fluids in a wide range of parameters. 

In this work, we consider a homogeneous and isotropic quantum cosmological system (universe) described by the Robertson--Walker metric and initially filled 
with a uniform scalar field. The potential energy of the scalar field is taken in the form of a sum of power-law terms. As we will see, each term in the sum can be 
connected to an era of dominance of one or another matter component. The well-known conditions for the conservation of the energy-momentum tensor for 
the scalar field (the Klein--Gordon equation) will not be discussed here. The proposed approach is based on quantum geometrodynamics for the maximally 
symmetric space. In the framework of quantum theory, the differential equations for the scalar field, whose potential energy is defined by one term from the 
sum, are obtained. The solutions to these equations are found in analytical form. Wave functions of the universe are calculated as exact analytical solutions to 
the quantum Hamiltonian constraint (Wheeler--DeWitt) equation. For this purpose, the scalar field is not taken as such, but, depending on the form of potential 
energy, is effectively reduced to one or another type of barotropic matter. In other words, the universe is viewed as dominated by one form or another of 
barotropic matter. Boundary conditions are not imposed, and the solution to the equations is sought in the most general form.

Since it is convenient to deal with dimensionless variables, we use the modified Planck system of units throughout the paper. 
The $l_{p} = \sqrt{2 G \hbar/(3 \pi c^{3})}$ is taken as a unit of length, the $\rho_{p} = 3 c^{4}/(8 \pi G l_{p}^{2})$ is a unit of energy density, 
$\phi_{p} = \sqrt{3 c^{4}/ (8 \pi G)}$ is a unit of scalar field and so on.

\section{General description}\label{sec:2}
\subsection{Classical theory}\label{sec:2.1}
We will consider the cosmological system (universe) originally filled with matter described by the scalar field. The general action can be written as
\begin{equation}\label{1}
S = S_{EH} + S_{PF},
\end{equation}
where
\begin{equation}\label{2}
S_{EH} = - \frac{c^{3}}{16 \pi G} \int \! d^{4}x \sqrt{- g} R
\end{equation}
is the Einstein--Hilbert action for gravitational field,
\begin{equation}\label{3}
S_{PF} = \frac{1}{2} \int \! d^{4}x \sqrt{- g} \left[g^{\mu \nu} \phi_{, \mu} \phi_{, \nu}  - V(\phi) \right]
\end{equation}
is the action of the scalar field.

Assume that the universe is homogeneous and isotropic, and that the scalar field is uniform. The geometry of spacetime is described by the 
Robertson--Walker metric
\begin{equation}\label{4}
ds^{2} = a^{2} (\eta) \left[N^{2} d\eta^{2} - \frac{dr}{1 - \kappa r^{2}} + r^{2} \left(d \vartheta^{2} +\sin^{2} \vartheta d\varphi^{2}  \right) \right],
\end{equation}
where $a$ is the cosmic scale factor, $N$ is an arbitrary lapse function that specifies the time reference scale, $\eta$ is the ``arc time'', the parameter 
$\kappa = +1, 0, -1$ is the curvature parameter.

Using the ADM formalism \cite{ADM62} one can define a Hamiltonian $H$ in which all the terms have a clear physical meaning and reduce the action 
to the form
\begin{equation}\label{5}
S = \int \! d\eta \left\{\pi_{a} \frac{da}{d\eta} + \pi_{\phi} \frac{d\phi}{d\eta} - H \right\},
\end{equation}
where 
\begin{equation}\label{6}
H = \frac{N}{2} \left\{- \pi_{a}^{2} - \kappa a^{2}  + a^{4} \rho_{\phi} \right\}.
\end{equation}
The energy density of the scalar field $\phi$ with the potential $V(\phi)$ is given by
\begin{equation}\label{7}
\rho_{\phi} = \frac{2}{a^{6}} \pi_{\phi}^{2} + V(\phi)
\end{equation}
and the corresponding pressure is
\begin{equation}\label{8}
p_{\phi} = \frac{2}{a^{6}} \pi_{\phi}^{2} - V(\phi).
\end{equation}
It is easy to see that
\begin{equation}\label{9}
\rho_{\phi} + p_{\phi}  = \frac{4}{a^{6}} \pi_{\phi}^{2} \quad \mbox{and} \quad \rho_{\phi} - p_{\phi}  = 2 V(\phi).
\end{equation}
From the second relation, the equation of state of stiff matter, $\rho_{\phi} = p_{\phi}$, immediately follows, when the potential $V(\phi)$ vanishes, so that
$\rho_{\phi} \sim a^{-6}$.

The variation of the action with respect to the Lagrange multiplier $N$ leads to the Hamiltonian constraint equation
\begin{equation}\label{10}
- \pi_{a}^{2} - \kappa a^{2}  + a^{4} \rho_{\phi}  \approx 0.
\end{equation}

\subsection{Quantization}\label{sec:2.2}
In quantum theory, following the quantization approach proposed by Dirac \cite{Dir64,Dir50,Dir51}, the constraint is considered as a restriction on the allowed state vectors. Choosing a representation, in which all operators are functions of ``position'', one maps classical variables to corresponding operators,
\begin{equation}\label{11}
\begin{split}
a \mapsto \hat{a} & = a, \quad \pi_{a} \mapsto \hat{\pi}_{a} = - i \partial_{a}  \\
\phi \mapsto \hat{\phi}& = \phi, \quad \pi_{\phi} \mapsto \hat{\pi}_{\phi} = - i \partial_{\phi},
\end{split}
\end{equation}
satisfying the commutation relations
\begin{equation}\label{12}
[a, \hat{\pi}_{a}] = i, \quad [\phi, \hat{\pi}_{\phi}] = i, \quad [a, \phi] = [\hat{\pi}_{a}, \hat{\pi}_{\phi}] = [a, \hat{\pi}_{\phi}] = [\phi, \hat{\pi}_{a}] = 0.
\end{equation}
We can introduce the operators of the energy density and the pressure of the field $\phi$
\begin{equation}\label{13}
\hat{\rho}_{\phi} = - \frac{2}{a^{6}}\partial_{\phi}^{2} + V(\phi), \quad \hat{p}_{\phi} = - \frac{2}{a^{6}}\partial_{\phi}^{2} - V(\phi).
\end{equation}
The quantum Hamiltonian (Wheeler--DeWitt) equation takes the form
\begin{equation}\label{14}
\left(- \partial_{a}^{2} + \kappa a^{2}  - a^{4} \hat{\rho}_{\phi}  \right) \Psi (a, \phi) = 0.
\end{equation}
Define the observable energy density $\rho_{m}$ and the corresponding pressure $p_{m}$ of matter as the mean values of the energy density and pressure 
operators (\ref{13}) \cite{KK13},
\begin{equation}\label{15}
\rho_{m} = \left \langle u | \hat{\rho}_{\phi} | u \right \rangle, \quad p_{m} = \left \langle u | \hat{p}_{\phi} | u \right \rangle,
\end{equation}
where $u (x) = \langle x | u \rangle$ is the complete orthonormalized set of functions that is assumed to diagonalize the energy density and pressure operators 
and describes the scalar field in terms of some generalized variable $ x = x(a, \phi)$ (see below). By expanding the wave function $\Psi$ in a series over states $u (x)$,
\begin{equation}\label{16}
\Psi = \sumint u (x) \psi (a),
\end{equation}
where the state indices are omitted and summation over discrete and/or integration over continuous states $u (x)$ is implied, and by substituting it
into Eq.~(\ref{14}), we obtain the equation
\begin{equation}\label{17}
\left(- \partial_{a}^{2} + \kappa a^{2}  - a^{4} \rho_{m}  \right) \psi (a) = 0.
\end{equation}

\section{Scalar field model}\label{sec:3}
Taking the equations (\ref{15}) into account, we can represent matter in the universe explicitly as a cosmological fluid with the equation of state
\begin{equation}\label{18}
p_{m} = w \rho_{m},
\end{equation}
where 
\begin{equation}\label{19}
w = - 1 - \frac{a}{3} \frac{d \ln \rho_{m}}{d a}
\end{equation}
is the equation-of-state parameter.

Within the framework of classical cosmology, one can accept the view that during the evolution of the universe, 
the stiff matter domination era ($w = 1$), perfect gas domination era ($w = \frac{2}{3}$), 
radiation domination era ($w = \frac{1}{3}$), dust domination era ($w = 0$), cosmic strings domination era ($w = - \frac{1}{3}$), domain walls domination era 
($w = - \frac{2}{3}$), the era of the domination of cosmological constant in the form of de Sitter vacuum ($w = -1$), and phantom matter domination era 
($w < - 1$) follow one another in sequence. We will set aside the question of the order of succession of eras and their transition from one to another, as well as the question of their actual existence and observability.

We restrict ourselves to considering the potential energy of a scalar field in the form of a power law
\begin{equation}\label{20}
V (\phi) = P \sum_{\alpha}\lambda_{\alpha} \phi^{\alpha},
\end{equation}
where $\phi \in (- \infty, \infty)$, $\lambda_{\alpha}$ are the coupling constants, the index $\alpha$ can generally take any real value, whereas the ``projection operator'' $P$ selects 
only one term from the sum which describes a specific era. The particular form of the operator $P$ is irrelevant for the problem 
at hand. Note that it can be represented, for instance, as a product of (the Heaviside) step functions.

Next, we will examine each component of the sum (\ref{20}) separately, describing every particular era one by one.
It is convenient to make a scale transformation of the field $\phi$, and introduce a variable (a new field)
\begin{equation}\label{21}
x = \left(\frac{\lambda_{\alpha} a^{6}}{2} \right)^{\frac{1}{2 + \alpha}} \phi \quad (\mbox{for} \ \lambda_{\alpha} \neq 0).
\end{equation}
Then the operator $\hat{\rho}_{\phi}$ reduces to
\begin{equation}\label{22}
\hat{\rho}_{\phi} = 2 \left(\frac{\lambda_{\alpha}}{2} \right)^{\frac{2}{2 + \alpha}} a^{-\frac{6 \alpha}{2 + \alpha}} \left[- \partial_{x}^{2} + x^{\alpha} \right].
\end{equation}
The observable energy density of matter $\rho_{m}$ takes the form
\begin{equation}\label{23}
\rho_{m} = A a^{-\frac{6 \alpha}{2 + \alpha}}, \quad A \equiv  2 \epsilon \left(\frac{\lambda_{\alpha}}{2} \right)^{\frac{2}{2 + \alpha}},
\end{equation}
where the parameters $\epsilon$ are eigenvalues of the equation
\begin{equation}\label{24}
\left[- \partial_{x}^{2} + x^{\alpha} \right] u = \epsilon u.
\end{equation}
The equation-of-state parameter (\ref{19}) is
\begin{equation}\label{25}
w = \frac{\alpha - 2}{\alpha + 2},
\end{equation}
so that the scalar field turns into the effective barotropic fluid.

\subsection{Stiff matter}\label{sec:3.1}
Consider the model of a scalar field with extremely strong self-interaction, which could have occurred in the very early universe. This formally corresponds to 
the case $\alpha = \infty$. Then, from Eqs.~(\ref{23}) and (\ref{25}), it follows
\begin{equation}\label{26}
\rho_{m} = \frac{2 \epsilon}{a^{6}}, \quad w = 1.
\end{equation}
Such energy density and equation-of-state parameter corresponds to the stiff matter.
It was shown that the matter component with a stiff equation of state arises in the previously empty universe with nonvanishing spatial curvature in the 
semi-classical approximation to the quantum gravity equation \cite{KK25}.

On the other hand, a similar result can be obtained if the kinetic energy of the scalar field significantly exceeds the potential energy.
For a free scalar field ($V(\phi) = 0$), the wave function satisfies the equation
\begin{equation}\label{27}
- \partial_{\phi}^{2} u = \epsilon u.
\end{equation}
It is convenient to impose the condition of periodicity
\begin{equation}\label{28}
u (\phi_{0}) = u (- \phi_{0}),
\end{equation}
where $\phi_{0}$ can be chosen sufficiently large so that, in the infinite limit, the wave function describes a continuous spectrum state.
Then the energy density of matter is 
\begin{equation}\label{29}
\rho_{m} = \left \langle u \middle| - 2 a^{-6} \partial_{\phi}^{2} \middle| u \right \rangle = \frac{2 \epsilon}{a^{6}},
\end{equation}
where 
\begin{equation}\label{30}
u (\phi) = \frac{1}{\sqrt{2 \phi_{0}}}\, e^{i \sqrt{2 \epsilon} \phi}
\end{equation}
is the wave function of a free scalar field normalized to unity in a one-dimensional ``box'' of length $2 \phi_{0}$.

The general solutions of Eq.~(\ref{17}) for the wave function $\psi (a)$ with the matter energy density (\ref{26}) can be obtained. For spatially flat universe 
($\kappa = 0$), we get
\begin{equation}\label{31}
\begin{split}
\psi (a) & = C_{1} a^{\frac{1 + \sqrt{1 - 4 A}}{2}} + C_{2} a^{\frac{1 - \sqrt{1 - 4 A}}{2}}, \quad \mbox{if} \ A < \frac{1}{4}, \\
\psi (a) & = \sqrt{a}\left [C_{1} \cos (\frac{1}{2}\sqrt{4A -1} \ln a) + C_{2} \sin (\frac{1}{2}\sqrt{4A -1} \ln a) \right], \quad \mbox{if} \ A > \frac{1}{4}.
\end{split}
\end{equation}
Here and below $C_{1}$ and $C_{2}$ are constants, and $A \equiv 2 \epsilon$.

For spatially closed universe ($\kappa = 1$), the general solution is
\begin{equation}\label{32}
\psi (a) = \left(\frac{a^{2}}{2}\right)^{\frac{1}{4}} \left[C_{1} I_{\nu}\left(\frac{a^{2}}{2}\right) +C_{2} K_{\nu}\left(\frac{a^{2}}{2}\right) \right], \quad 
\nu = \frac{1}{4} \sqrt{1 - 4 A},
\end{equation}
where $I_{\nu}(x)$ and $K_{\nu}(x)$ are modified Bessel functions of the first kind and second kind, respectively. The solution can be written equivalently 
in terms of Bessel functions with imaginary argument. For $A > \frac{1}{4}$, the order $\nu$ becomes imaginary, and the solutions are expressed in terms of 
modified Bessel functions of imaginary order.

For spatially open universe ($\kappa = - 1$), we get
\begin{equation}\label{33}
\psi (a) = \left(\frac{a^{2}}{2}\right)^{\frac{1}{4}} \left[C_{1} J_{\nu}\left(\frac{a^{2}}{2}\right) +C_{2} Y_{\nu}\left(\frac{a^{2}}{2}\right) \right], \quad 
\nu = \frac{1}{4} \sqrt{1 - 4 A},
\end{equation}
where $J_{\nu}(x)$ and $Y_{\nu}(x)$ are Bessel functions of the first kind and second kind, respectively. Equation (\ref{33}) still holds, if $A > \frac{1}{4}$ 
(Bessel functions of imaginary order).

\subsection{Perfect gas}\label{sec:3.2}
Examine the model of a scalar field that produces a perfect gas with $w = \frac{2}{3}$. According to Eq.~(\ref{25}), it corresponds to the parameter 
$\alpha = 10$. The energy density of matter (\ref{23}) reduces to
\begin{equation}\label{34}
\rho_{m} = \frac{A}{a^{5}}, \quad A \equiv 2 \epsilon \left(\frac{\lambda_{10}}{2} \right)^{\frac{1}{6}}.
\end{equation}

Equation (\ref{24}) with $\alpha = 10$ has a form of a one-dimensional oscillator with a highly anharmonic potential. It does not reduce globally to  any of the 
classical special-function equations. Its solutions can be obtained, e.g., using the WKB method.
Away from the turning point $x_{0} = \epsilon^{1/10}$, in the classically allowed region, where $Q \equiv \epsilon - x^{10} > 0$, we have
\begin{equation}\label{341}
u (x) \approx \frac{1}{Q(x)^{1/4}}\left[C_{1} \cos \left(\int_{x_{0}}^{x}\!dx \sqrt{Q(x)} - \frac{\pi}{4}\right) + 
C_{2} \sin \left(\int_{x_{0}}^{x}\!dx \sqrt{Q(x)} - \frac{\pi}{4}\right) \right],
\end{equation}
while in the classically forbidden region, $Q \equiv \epsilon - x^{10} < 0$, the solution can be written as
\begin{equation}\label{342}
u (x) \approx \frac{1}{|Q(x)|^{1/4}}\left[C_{1} \exp \left(- \int_{x_{0}}^{x}\!dx \sqrt{|Q(x)|} \right) + 
C_{2} \exp \left(\int_{x_{0}}^{x}\!dx \sqrt{|Q(x)|} \right)  \right].
\end{equation}
The integrals in Eqs.~(\ref{341}) and (\ref{342}) can be expressed in terms of incomplete Beta function
\begin{equation}\label{343}
\int_{x_{0}}^{x}\!dx \sqrt{\epsilon - x^{10}} = \frac{\epsilon^{3/5}}{10} \left[B_{t} \left(\frac{1}{10}, \frac{3}{2} \right) - B_{1} \left(\frac{1}{10}, \frac{3}{2} \right) \right],
\ t = \frac{x^{10}}{10},
\end{equation}
\begin{equation}\label{344}
\int_{x_{0}}^{x}\!dx \sqrt{x^{10} - \epsilon} = - \frac{\epsilon^{3/5}}{10} \left[B_{t^{-1}} \left(- \frac{3}{5}, \frac{3}{2} \right) - B_{1} \left(- \frac{3}{5}, \frac{3}{2} \right) \right].
\end{equation}

The general solutions of Eq.~(\ref{17}) for the wave function $\psi (a)$ with the matter energy density (\ref{34}) have the form
\begin{equation}\label{35}
\psi (a) = 2 \sqrt{A a} \left[C_{1} J_{1} (2 \sqrt{A a}) + C_{2} Y_{1} (2 \sqrt{A a}) \right] \ \mbox{for} \ \kappa = 0.
\end{equation}
In the cases $\kappa \neq 0$, the equation (\ref{17}) with (\ref{34}) reduces to the canonical form of the biconfluent Heun equation 
\begin{equation}\label{351}
\frac{d^{2} y}{d z^{2}} + \left(\frac{1 + \upalpha}{z} - \upbeta - 2 z \right) \frac{d y}{d z} + 
\left(\upgamma - \upalpha - 2 - \frac{\updelta + (1 + \upalpha) \upbeta}{2 z} \right) y = 0,
\end{equation}
so that the general solution can be written as
\begin{equation}\label{36}
\psi (a) = e^{-\frac{a^{2}}{2}} \left[C_{1}\,  y_{1} (-1,0,0, -2A; a) + C_{2}\, a\,y_{2} (1,0,0, -2A; a) \right] \ \mbox{for} \ \kappa = 1,
\end{equation}
\begin{equation}\label{37}
\begin{split}
\psi (a) & = e^{i \frac{a^{2}}{2}} \left[C_{1}\, y_{1} (-1,0,0, -2A e^{\frac{i \pi}{4}}; e^{\frac{-i \pi}{4}} a) \right. \\
& + \left. C_{2}\, a\, y_{2} (1,0,0, -2A e^{\frac{i \pi}{4}}; e^{\frac{-i \pi}{4}} a) \right] \ \mbox{for} \ \kappa = - 1,
\end{split}
\end{equation}
where $y_{i} (\upalpha, \upbeta, \upgamma, \updelta ; z)$ are the solutions of the biconfluent Heun equation (\ref{351}) \cite{Ro95,SL00}.

\subsection{Radiation}\label{sec:3.3}
The radiation component with $w = \frac{1}{3}$ is generated by the scalar field potential term with $\alpha = 4$ in Eq.~(\ref{20}).
The energy density (\ref{23}) for radiation takes the form
\begin{equation}\label{38}
\rho_{m} = \frac{A}{a^{4}}, \quad A \equiv 2 \epsilon \left(\frac{\lambda_{4}}{2} \right)^{\frac{1}{3}}.
\end{equation}

Equation (\ref{24}) with $\alpha = 4$ can be reduced to the form of the triconfluent Heun equation 
\begin{equation}\label{381}
\frac{d^{2} y}{d z^{2}} + \left(\upgamma + z \right) z \frac{d y}{d z} + 
\left(\upalpha z - q \right) y = 0,
\end{equation}
The general solution for a rescaled scalar field $x$ can be represented as 
\begin{equation}\label{39}
u (x) = C_{1}\, e^{\frac{x^{3}}{3}} y_{1} (1,0, - 2^{-2/3} \epsilon; 2^{1/3} x) + 
C_{2}\, e^{- \frac{x^{3}}{3}} y_{2} (1,0, - 2^{-2/3} \epsilon; - 2^{1/3} x),
\end{equation}
where $y_{i} (\upalpha, \upgamma, q ; z)$ are two linearly independent solutions of the triconfluent Heun equation (\ref{381}) \cite{Ro95,SL00}.

The general solutions of Eq.~(\ref{17}) for the wave function $\psi (a)$ with the matter energy density (\ref{38}) are
\begin{equation}\label{40}
\psi (a) = C_{1} \cos (\sqrt{A} a) + C_{2} \sin (\sqrt{A} a) \ \mbox{for} \ \kappa = 0,
\end{equation}
\begin{equation}\label{41}
\psi (a) = C_{1} e^{-\frac{a^{2}}{2}} H_{n} (a) + C_{2}\, 2^{- \frac{n + 1}{2}} e^{\frac{a^{2}}{2}} U\left( \frac{n + 1}{2}, \frac{1}{2}, - a^{2} \right), \ A = 2n + 1, 
\ \mbox{for} \ \kappa = 1,
\end{equation}
\begin{equation}\label{42}
\psi (a) = C_{1} D_{\nu} \left((1 + i ) a\right) + C_{2} D_{\nu} \left(-(1 + i ) a\right), \ \nu = - \frac{1}{2} - \frac{i A}{2}, \ \mbox{for} \ \kappa = - 1,
\end{equation}
where $H_{n}(x)$ is the Hermite polynomial of degree $n$, $U(a,b,z)$ is the confluent hypergeometric function of the second kind, 
$D_{\nu}(x)$ is the parabolic-cylinder function.

\subsection{Dust}\label{sec:3.4}
The scalar field potential term with $\alpha = 2$ in Eq.~(\ref{20}) produces the dust matter component with $w = 0$.
The energy density (\ref{23}) for dust becomes
\begin{equation}\label{43}
\rho_{m} = \frac{A}{a^{3}}, \quad A \equiv 2 \epsilon \left(\frac{\lambda_{2}}{2} \right)^{\frac{1}{2}}.
\end{equation}

The general solution of Eq.~(\ref{24}) with $\alpha = 2$ for a redefined field $x$ can be represented as 
\begin{equation}\label{44}
u (x) = C_{1} e^{-\frac{x^{2}}{2}} H_{n} (x) + C_{2}\, 2^{- \frac{n + 1}{2}} e^{\frac{x^{2}}{2}} U\left( \frac{n + 1}{2}, \frac{1}{2}, - x^{2} \right), \ \epsilon = 2n + 1.
\end{equation}

The general solutions of Eq.~(\ref{17}) for the wave function $\psi (a)$ with the matter energy density (\ref{43}) have the form
\begin{equation}\label{45}
\psi (a) = C_{1} \mbox{Ai} (- A^{-1/3} a) + C_{2} \mbox{Bi} (- A^{-1/3} a) \ \mbox{for} \ \kappa = 0,
\end{equation}
where $\mbox{Ai} (x)$ and $\mbox{Bi}(x)$ are the Airy function of the first and second kind, respectively,
\begin{equation}\label{46}
\psi (a) = C_{1} D_{\nu} \left(\sqrt{2} \left(a - \frac{A}{2} \right) \right) + C_{2} D_{\nu} \left(- \sqrt{2} \left(a - \frac{A}{2} \right) \right), \ 
\nu = - \frac{1}{2} + \frac{A^{2}}{8}, \ \mbox{for} \ \kappa = 1,
\end{equation}
\begin{equation}\label{47}
\psi (a) = C_{1} D_{\nu} \left(\sqrt{2}e^{i \pi / 4} \left(a + \frac{A}{2} \right) \right) + C_{2} D_{\nu} \left(- \sqrt{2}e^{i \pi / 4} \left(a + \frac{A}{2} \right) \right), 
\ \nu = - \frac{1}{2} - \frac{i A^{2}}{8}, \ \mbox{for} \ \kappa = - 1.
\end{equation}

\subsection{Cosmic strings}\label{sec:3.5}
The cosmic strings with $w = - \frac{1}{3}$ is brought about by the scalar field potential term with $\alpha = 1$ in Eq.~(\ref{20}). The energy density (\ref{23}) for 
the cosmic strings is
\begin{equation}\label{48}
\rho_{m} = \frac{A}{a^{2}}, \quad A \equiv 2 \epsilon \left(\frac{\lambda_{1}}{2} \right)^{\frac{2}{3}}.
\end{equation}
Equation (\ref{24}) with $\alpha = 1$ for a rescaled field $x$ has the following general solution
\begin{equation}\label{49}
u (x) = C_{1} \mbox{Ai} (x - \epsilon) + C_{2} \mbox{Bi} (x - \epsilon).
\end{equation}
The general solutions of Eq.~(\ref{17}) with the matter energy density (\ref{48}) are
\begin{equation}\label{50}
\psi (a) = \sqrt{a} \left[C_{1} J_{1/4} \left(\frac{\sqrt{A - \kappa}}{2} a^{2} \right) + C_{2} Y_{1/4} \left(\frac{\sqrt{A - \kappa}}{2} a^{2} \right) \right] \ \mbox{for} \ \kappa = 0, 1, -1.
\end{equation}

\subsection{Domain walls}\label{sec:3.6}
The domain walls with $w = - \frac{2}{3}$ can be obtained when considering the scalar field potential term with $\alpha = \frac{2}{5}$ in Eq.~(\ref{20}). 
The energy density (\ref{23}) for the domain walls has the form
\begin{equation}\label{51}
\rho_{m} = \frac{A}{a}, \quad A \equiv 2 \epsilon \left(\frac{\lambda_{2/5}}{2} \right)^{\frac{5}{6}}.
\end{equation}

Equation (\ref{24}) with $\alpha = 2/5$ does not reduce to a commonly known special-function equation. The solutions can be obtained using the WKB 
method. They have the form (\ref{341}) for $Q =  \epsilon - x^{2/5} > 0$ and (\ref{342}) for $Q < 0$.
A Puiseux series expansion can be used to obtain the solutions for small $x$.

The integrals in Eqs.~(\ref{341}) and (\ref{342}) for $Q =  \epsilon - x^{2/5}$ can be written as
\begin{equation}\label{511}
\begin{split}
& \int_{x_{0}}^{x}\!dx \sqrt{\epsilon - x^{2/5}} \\
& = \frac{5}{48} \left[x^{1/5} \sqrt{\epsilon - x^{2/5}} (8 x^{4/5} - 2 \epsilon x^{2/5} - 3 \epsilon^{2}) + 
3 \epsilon^{3} \tan^{-1} \left(\frac{x^{1/5}}{\sqrt{\epsilon - x^{2/5}} } \right) - 3 \epsilon^{3} \frac{\pi}{2} \right],
\end{split}
\end{equation}
\begin{equation}\label{512}
\begin{split}
& \int_{x_{0}}^{x}\!dx \sqrt{x^{2/5} - \epsilon} \\
& = \frac{5}{48} \left[x^{1/5} \sqrt{x^{2/5} - \epsilon} (8 x^{4/5} - 2 \epsilon x^{2/5} - 3 \epsilon^{2}) - 
3 \epsilon^{3} \tanh^{-1} \left(\frac{x^{1/5}}{\sqrt{x^{2/5} - \epsilon} } \right) - 3 \epsilon^{3} i \frac{\pi}{2} \right],
\end{split}
\end{equation}

The general solutions of Eq.~(\ref{17}) with the matter energy density (\ref{51}) can be written in the compact form for $\kappa = 0$,
\begin{equation}\label{52}
\psi (a) = \sqrt{a} \left[C_{1} J_{1/5} \left(\frac{2}{5} \sqrt{A} a^{5/2} \right) + C_{2} Y_{1/5} \left(\frac{2}{5} \sqrt{A} a^{5/2} \right) \right].
\end{equation}
For $\kappa = 1$ or $\kappa = - 1$, Eq.~(\ref{17}) cannot be exactly reduced to a standard special-function equation. The WKB method or a local power-series 
expansion near $a = 0$ can be used to analyze it. Away from the turning points, the WKB solutions are
\begin{equation}\label{521}
\begin{split}
\psi (a) & \approx \frac{1}{(A a^{3} - a^{2})^{1/4}} \left\{C_{1} \cos \left[ \frac{2}{A^{2}} \left(\frac{1}{3} (A a - 1)^{3/2} + 
\frac{1}{5} (A a - 1)^{5/2} \right) - \frac{\pi}{4}\right] \right. \\
& + \left. C_{2} \sin \left[ \frac{2}{A^{2}} \left(\frac{1}{3} (A a - 1)^{3/2} + 
\frac{1}{5} (A a - 1)^{5/2} \right) - \frac{\pi}{4}\right] \right\}, \ a > \frac{1}{A};
\end{split}
\end{equation}
\begin{equation}\label{522}
\begin{split}
\psi (a) & \approx \frac{1}{(a^{2} - A a^{3})^{1/4}} \left\{C_{1} \exp \left[ \frac{2}{A^{2}} \left(\frac{1}{3} (1 - A a)^{3/2} - 
\frac{1}{5} (1 - A a)^{5/2} \right) \right] \right. \\
& + \left. C_{2} \exp \left[ - \frac{2}{A^{2}} \left(\frac{1}{3} (1 - A a)^{3/2} - 
\frac{1}{5} (1 - A a)^{5/2} \right) \right] \right\}, \ a < \frac{1}{A},  \ \mbox{for} \ \kappa = 1,
\end{split}
\end{equation}
\begin{equation}\label{523}
\begin{split}
\psi (a) & \approx \frac{1}{(A a^{3} + a^{2})^{1/4}} \left\{C_{1} \cos \left[ \frac{2}{A^{2}} \left(- \frac{1}{3} (A a + 1)^{3/2} + 
\frac{1}{5} (A a + 1)^{5/2} \right) \right] \right. \\
& + \left. C_{2} \sin \left[ \frac{2}{A^{2}} \left(- \frac{1}{3} (A a + 1)^{3/2} + 
\frac{1}{5} (A a + 1)^{5/2} \right) \right] \right\},  \ \mbox{for} \ \kappa = -1.
\end{split}
\end{equation}

\subsection{de Sitter vacuum}\label{sec:3.7}
The de Sitter vacuum with $w = -1$ is reproduced by the scalar field potential term with $\alpha = 0$ in Eq.~(\ref{20}).
The energy density (\ref{23}) takes the form
\begin{equation}\label{53}
\rho_{m} = A = \frac{\Lambda}{3} = const, \quad A \equiv \epsilon \lambda_{0},
\end{equation}
where $\Lambda$ is the cosmological constant.

The general solution of Eq.~(\ref{24}) with $\alpha = 0$ is trivial,
\begin{equation}\label{54}
\begin{split}
u (x) & = C_{1} e^{\sqrt{1 - \epsilon} x} + C_{2} e^{- \sqrt{1 - \epsilon} x}, \ \mbox{for} \ \epsilon < 1, \\
u (x) & = C_{1} \cos (\sqrt{\epsilon - 1} x) + C_{2} \sin (\sqrt{\epsilon - 1} x), \ \mbox{for} \ \epsilon > 1.
\end{split}
\end{equation}

The general solutions of Eq.~(\ref{17}) with the matter energy density (\ref{53}) are as follows. 
For $\kappa = 0$, we have
\begin{equation}\label{55}
\psi (a) = \sqrt{a} \left[C_{1} J_{1/6} \left(\frac{\sqrt{A}}{3} a^{3} \right) + C_{2} Y_{1/6} \left(\frac{\sqrt{A}}{3} a^{3} \right) \right].
\end{equation}
For $\kappa = 1$, the solution is
\begin{equation}\label{56}
\psi (a) = C_{1}\, e^{\alpha \frac{a^{3}}{3} + \frac{a}{2 \alpha}} y_{1} (1, \upgamma, q_{1}; z) + C_{2}\, e^{- \alpha \frac{a^{3}}{3} - \frac{a}{2 \alpha}} 
y_{2} (1, - \upgamma, q_{2}; - z) 
\end{equation}
with
\begin{equation*} 
\alpha = i \sqrt{A}, \ z = s a + t, \ s^{3} = 2 \alpha, \ t^{2} = - \frac{1}{\alpha s}, \ \upgamma = - 2 t, \ 
q_{1} = t - \frac{1}{4} t^{4}, \ q_{2} = - t - \frac{1}{4} t^{4}. 
\end{equation*}
For $\kappa = - 1$, we get
\begin{equation}\label{57}
\psi (a) = C_{1}\, e^{- \alpha \frac{a^{3}}{3} + \beta_{1} \lambda_{1} a} y_{1} (1, -2, q_{1}; k \lambda_{1} a + 1) + 
C_{2}\, e^{\alpha \frac{a^{3}}{3} + \beta_{2} \lambda_{2} a} y_{2} (1, -2, q_{2}; k \lambda_{2} a + 1)
\end{equation}
with
\begin{equation*}
\alpha = i \sqrt{A}, \ \lambda_{j}^{3} = (-1)^{j-1} \frac{2}{3} \alpha, \ \beta_{j} = \frac{1}{3 \lambda_{j}}, q_{j} = - \frac{1}{9 \lambda_{j}^{8} k^{2}} + 1, \ 
k^{3} = - 3. 
\end{equation*}
The functions $y_{i} (\upalpha, \upgamma, q ; z)$ in Eqs.~(\ref{56}) and (\ref{57}) are the solutions of the triconfluent Heun equation (\ref{381}).

\subsection{Phantom matter}\label{sec:3.8}
Consider a special case of the phantom matter with $w  = - \frac{4}{3}$ as an illustration. Such a matter component is described by the scalar field potential 
term with $\alpha = - \frac{2}{7}$ in Eq.~(\ref{20}). The energy density (\ref{23}) for the phantom matter has the form
\begin{equation}\label{58}
\rho_{m} = A a, \quad A \equiv 2 \epsilon \left(\frac{\lambda_{-2/7}}{2} \right)^{\frac{7}{6}}.
\end{equation}

The solution of Eq.(\ref{24}) with $\alpha = - 2/7$ can be obtained using the WKB method. They have the form (\ref{341}) for $Q =  \epsilon - x^{-2/7} > 0$ 
and (\ref{342}) for $Q < 0$. The integrals in Eqs.~(\ref{341}) and (\ref{342}) for $Q =  \epsilon - x^{-2/7}$ can be calculated explicitly,
\begin{equation}\label{59}
\int_{x_{0}}^{x}\!dx \sqrt{\epsilon - x^{-2/7}} = \frac{1}{\epsilon^{3}} \left(\epsilon - x^{-2/7} \right)^{3/2} \left[\epsilon^{2} x^{4/7} + \frac{4}{5} \epsilon x^{2/7}
+ \frac{8}{15} \right],
\end{equation}
\begin{equation}\label{60}
\int_{x_{0}}^{x}\!dx \sqrt{x^{-2/7} - \epsilon} = - \frac{1}{\epsilon^{3}} \left(x^{-2/7} - \epsilon \right)^{3/2} \left[\epsilon^{2} x^{4/7} + \frac{4}{5} \epsilon x^{2/7}
+ \frac{8}{15} \right].
\end{equation}

For $\kappa = 0$, the general solution of Eq.~(\ref{17}) with the matter energy density (\ref{58}) is
\begin{equation}\label{61}
\psi (a) = \sqrt{a} \left[C_{1} J_{1/7} \left(\frac{2 \sqrt{A}}{7} a^{7/2} \right) + C_{2} Y_{1/7} \left(\frac{2\sqrt{A}}{7} a^{7/2} \right) \right].
\end{equation}
For $\kappa \neq 0$, the WKB solutions away from the turning points can be written in terms of incomplete Beta functions
\begin{equation}\label{62}
\begin{split}
\psi (a) & \approx \frac{1}{(A a^{5} - a^{2})^{1/4}} \left\{C_{1} \cos \left[ \frac{1}{3 A^{2/3}} B_{t} \left(\frac{3}{2}, - \frac{7}{6} \right) - \frac{\pi}{4}\right] \right. \\
& + \left. C_{2} \sin \left[ \frac{1}{3 A^{2/3}} B_{t} \left(\frac{3}{2}, - \frac{7}{6} \right) - \frac{\pi}{4}\right], \right\}, \ t = \frac{A a^{3} - 1}{A a^{3}}, \ a > \frac{1}{A^{1/3}};
\end{split}
\end{equation}
\begin{equation}\label{63}
\begin{split}
\psi (a) & \approx \frac{1}{(a^{2} - A a^{5})^{1/4}} \left\{C_{1} \exp \left[ - \frac{1}{3A^{2/3}} \left(B_{s} \left(\frac{2}{3}, \frac{3}{2} \right) -
B_{1} \left(\frac{2}{3}, \frac{3}{2} \right) \right) \right] \right. \\
& + \left. C_{2}\exp \left[ \frac{1}{3A^{2/3}} \left(B_{s} \left(\frac{2}{3}, \frac{3}{2} \right) -
B_{1} \left(\frac{2}{3}, \frac{3}{2} \right) \right) \right] \right\}, \ s = A a^{3}, \ a < \frac{1}{A^{1/3}},  \ \mbox{for} \ \kappa = 1,
\end{split}
\end{equation}
\begin{equation}\label{64}
\begin{split}
\psi (a) & \approx \frac{1}{(A a^{5} + a^{2})^{1/4}} \left\{C_{1} \cos \left[ \frac{1}{3 A^{2/3}} B_{r} \left(\frac{2}{3}, - \frac{7}{6} \right) \right] \right. \\
& + \left. C_{2} \sin \left[ \frac{1}{3 A^{2/3}} B_{r} \left(\frac{2}{3}, - \frac{7}{6} \right) \right] \right\}, \ r = \frac{A a^{3}}{A a^{3} + 1}, \ \mbox{for} \ \kappa = -1.
\end{split}
\end{equation}

\section{Discussion}\label{sec:4}
This work is focused on solving equations of quantum geometrodynamics for the maximally symmetric space. Explicit expressions for the wave function of the 
universe, in which one of the possible (including theoretical) components of matter-energy dominates, are calculated. These components are generated by a 
primordial scalar field, for which quantum equations of motion are also obtained and their solutions are found in each individual case. The wave functions of the 
universe are represented as two functions, one of which decreases with the growth of the cosmic scale factor $a$, and the other tends to infinity. We leave the 
question of boundary conditions for the wave function aside, since it is directly related to the choice of a specific cosmological system, which we do not 
concretize here.

Solutions to the equation (\ref{17}) for the wave function $\psi (a)$, are formally obtained for all values $a > 0$. However, since the density of matter-energy 
$\rho_{m}$ (\ref{23}) varies depending on the era in certain regions of the variable $a$, where only one of the energy density components dominates, the wave 
function should only be considered on restricted intervals of $a$. A smooth transition from one region to another is provided by merging the logarithmic 
derivatives of the wave functions at their boundaries. We do not give the corresponding formulas in this work. Note that the coupling constant $A$ (\ref{23}) of 
the energy density is determined by the coupling constants $\lambda_{\alpha}$ of the scalar field and their energy states $\epsilon$.

In conclusion, we remark that the question of applying quantum mechanics to our Universe as a whole was intensively discussed in the literature of the 1980s 
and 1990s \cite{HH83,V86,HH90,GH90,BH98}. It is generally accepted that the Universe in the Planck era should be described as a quantum system. The 
transition to general relativity can be viewed as a sequential transition from quantum theory to quasi-classical theory and then to classical theory. However, it 
has been suggested that the Universe may retain certain quantum properties at later times \cite{Ki08}. For example, the cosmological constant in the form of 
quintessence allows us to reconcile the theoretical value of $\Lambda$ obtained within the framework of quantum field theory with observations indicating that 
the density of dark energy is close in magnitude to the energy density of matter in the modern era. It was argued that since inflation takes place at scales that 
are a few orders of magnitude below the Planck scale, quantum gravitational effects may come into play during that stage \cite{KTV19}. The same applies to 
the early post-inflation era \cite{KK25}. The universe as a whole is quantum in nature, but appears classical in most circumstances. However, there may be 
situations where its quantum nature reveals itself, not only at Planck scales. Therefore, studying the properties of wave functions for different types of matter 
content seems to be important and interesting. At the same time, the role played by the wave function may not be limited to understanding the initial conditions 
of the evolution of the universe.

\section*{Acknowledgements}
This work was partially supported by The National Academy of
Sciences of Ukraine (Projects No.~0121U109612 and  No.~0122U000886) and by a grant from Simons Foundation International SFI-PD-Ukraine-00014573, PI LB.

\end{document}